\documentclass[prb,twocolumn,showpacs,preprintnumbers]{revtex4}
\usepackage{graphics}
\usepackage{graphicx}
\usepackage{dcolumn} 
\usepackage{bm}
\usepackage{float}
\usepackage{epsfig}
\begin{document}
\title{Destruction of the small Fermi surfaces in Na$_x$CoO$_2$ by Na disorder}
\author{D.J. Singh$^1$ and Deepa Kasinathan$^2$}
\affiliation{$^1$Materials Science and Technology Division,
Oak Ridge National Laboratory, Oak Ridge,TN 37831-6032}
\affiliation{$^2$Department of Physics, University of California Davis,
  Davis, CA 95616}
\date{\today}

\begin{abstract}
We show using density functional calculations that the small $e_g'$ 
Fermi surfaces in Na$_x$CoO$_2$
are destroyed by Na disorder. This provides a means to resolve the
prediction of these sections in band structure calculations with their
non-observation in angle resolved photoemission experiments.
\end{abstract}

\maketitle

The layered oxide, Na$_x$CoO$_2$, has attracted considerable interest
because of its enhanced thermoelectric properties, \cite{terasaki}
and because, when hydrated the material becomes a superconductor,
possibly with unconventional pairing.
\cite{takada,mazin}.
Na$_x$CoO$_2$ consists of triangular
sheets of Co ions, with nominal $d$ electron count,
5+$x$. These are coordinated by distorted edge sharing octahedra, 
made from triangular sheets of O ions above and below the Co sheets.
These CoO$_2$ tri-layers are stacked in an alternating fashion to 
form prismatic sites, which contain the Na.

Establishing the
electronic structure is requisite for understanding of both the
thermoelectric behavior and the superconductivity.
Local density approximation (LDA) band structure calculations,
\cite{singh2000,johannes}
show a Fermi level that lies near the top of a narrow manifold of $t_{2g}$
bands, which is separated by a gap from the higher lying $e_g$ derived states.
With the actual rhombohedral Co site symmetry, the three $t_{2g}$ orbitals
are further divided into an $a_g$ and two degenerate $e_g$ symmetry
orbitals, denoted $e_g'$ to distinguish them from the higher lying $e_g$
manifold defined by the primary crystal field splitting.
Significantly, local density approximation band structure calculations
show the presence of two sheets of Fermi surface over a wide range of $x$.
These are a large $a_g$ derived hole cylinder around $\Gamma$ and
six small hole sections along the hexagonal $\Gamma$ - $K$ lines.
The calculated Fermi surfaces of the hydrated
superconducting compound similarly show both $a_g$ and small $e_g'$
sections. \cite{johannes-h2o}
While these $e_g'$ sections are small, and therefore are not expected
to contribute substantially to the conduction, they contribute strongly
to the density of states and should be very important for
scattering, magnetic susceptibility and other properties because of
the two dimensional nature of the material and their heavy mass.
\cite{lee,johannes-nest}

In contrast to the band structure calculations, angle resolved photoemission
(ARPES) experiments by several
groups at various doping levels
observe the large $a_g$ cylinder but do not find the small $e_g'$ sections.
\cite{hasan,yang,yang-2,qian,hasan2}
Early on, \cite{singh2000} it was noted that although metallic,
Na$_x$CoO$_2$ had a small band width compared to plausible values
for the on-site Coulomb repulsion, $U$. One explanation that has been
advanced for the absence of the $e_g'$ sections is that they are
destroyed by correlation effects. LDA+U calculations, which incorporate
Coulomb repulsion in a static mean field like way, have indeed shown
that the small sections are removed for reasonable values of $U$.
\cite{lee,zhang}
However, the LDA+U approximation also favors insulating, charge ordered
ground states. \cite{lee}
Removal of the small sections is also found in the large $U$ limit
Gutzwiller approximation. \cite{zhou}
However, in metals, fluctuations, including charge and orbital fluctuations
are favored by kinetic energy considerations and the presence of $e_g'$
sections would open more degrees of freedom for fluctuations.
Dynamical mean field calculations, which incorporate physics like the LDA+U
approximation, but in a beyond mean field manner that incorporates fluctuations,
yielded the opposite tendency to the LDA+U approximation, {\em i.e}
an enlargement of the $e_g'$ pockets over the LDA. \cite{ishida}
Low temperature specific heat measurements on superconducting
hydrated Na$_x$CoO$_2$ show two distinct energy gaps,
\cite{oeschler}
which is difficult to
understand if there is only one, simple Fermi surface.
Thus, the discrepancy between the ARPES and LDA Fermi surfaces remains
a puzzle. Here we resolve this puzzle by showing that the small
sections are destroyed by scattering due to the Na disorder inherent
in the structural chemistry of Na$_x$CoO$_2$.

As mentioned, the Na ions are contained in O prisms defined by the
O triangular lattice. There are two types of sites: Those directly
above a Co ion, denoted Na$_{\rm Co}$ here, and those above the
holes in the Co sheet, denoted Na$_h$
(these are directly above the O atoms in the
O layer on the opposite side of the Co sheet). There is one
Na$_{\rm Co}$ and one Na$_h$ site per Co, with a total Na filling
of $x$ per Co, or $x/2$ per site.
However, the number of available sites is smaller since occupation
of an Na$_{\rm Co}$ next to an occupied Na$_h$ site or vice versa
would be highly unfavorable due to the short Na - Na distance of 
1.64 \AA$~$ that would result.
In fact, local structures that within a given Na
layer have Na of a given type (Na$_{\rm Co}$ or Na$_h$) having
Na neighbors of the same type are favored. Also
Na in adjacent layers tend not to coordinate the same Co ion.
\cite{zand,zhang2,huang}

\begin{figure}[tbp]
\epsfig{width=0.46\columnwidth,angle=0,file=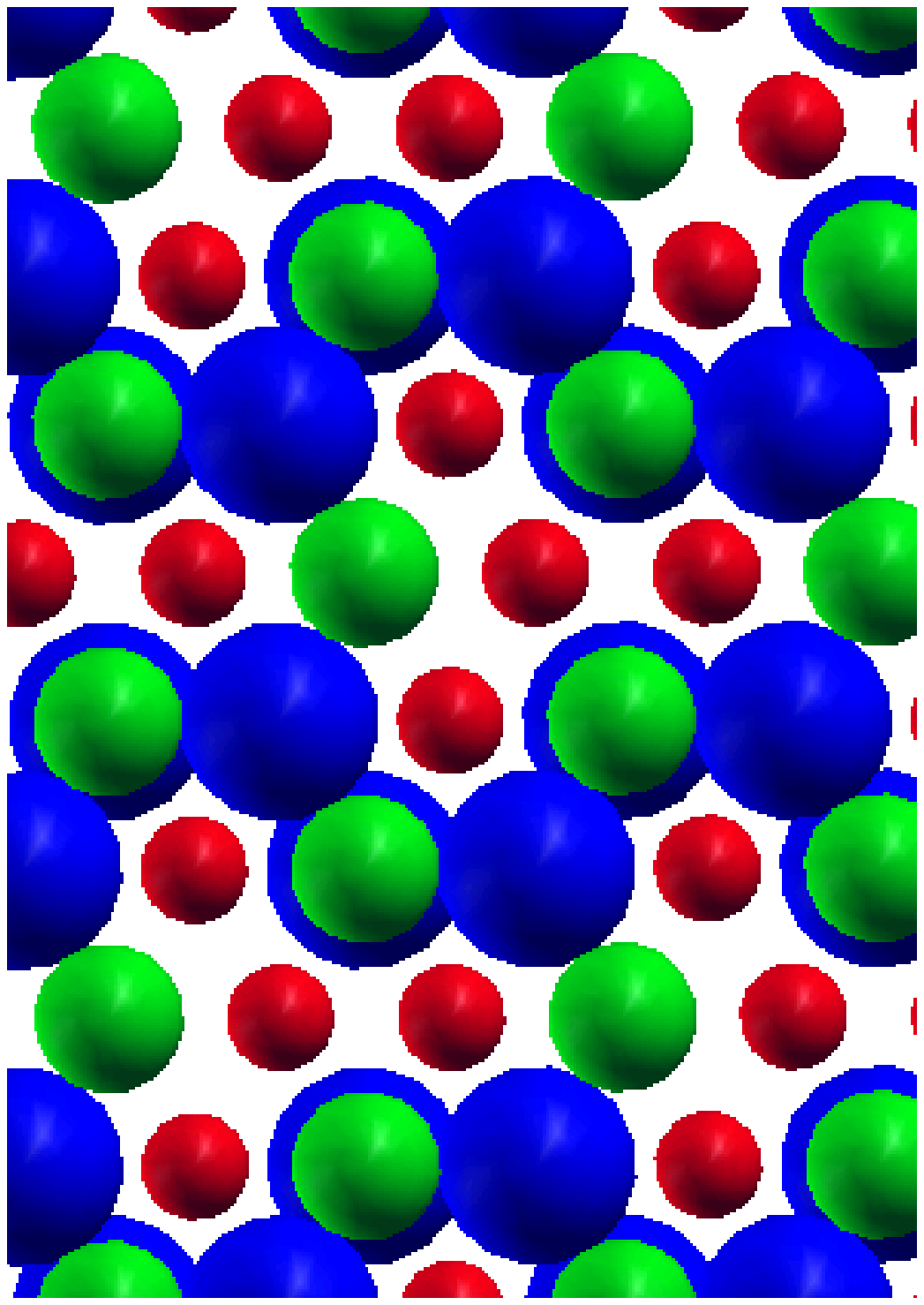}
\epsfig{width=0.46\columnwidth,angle=0,file=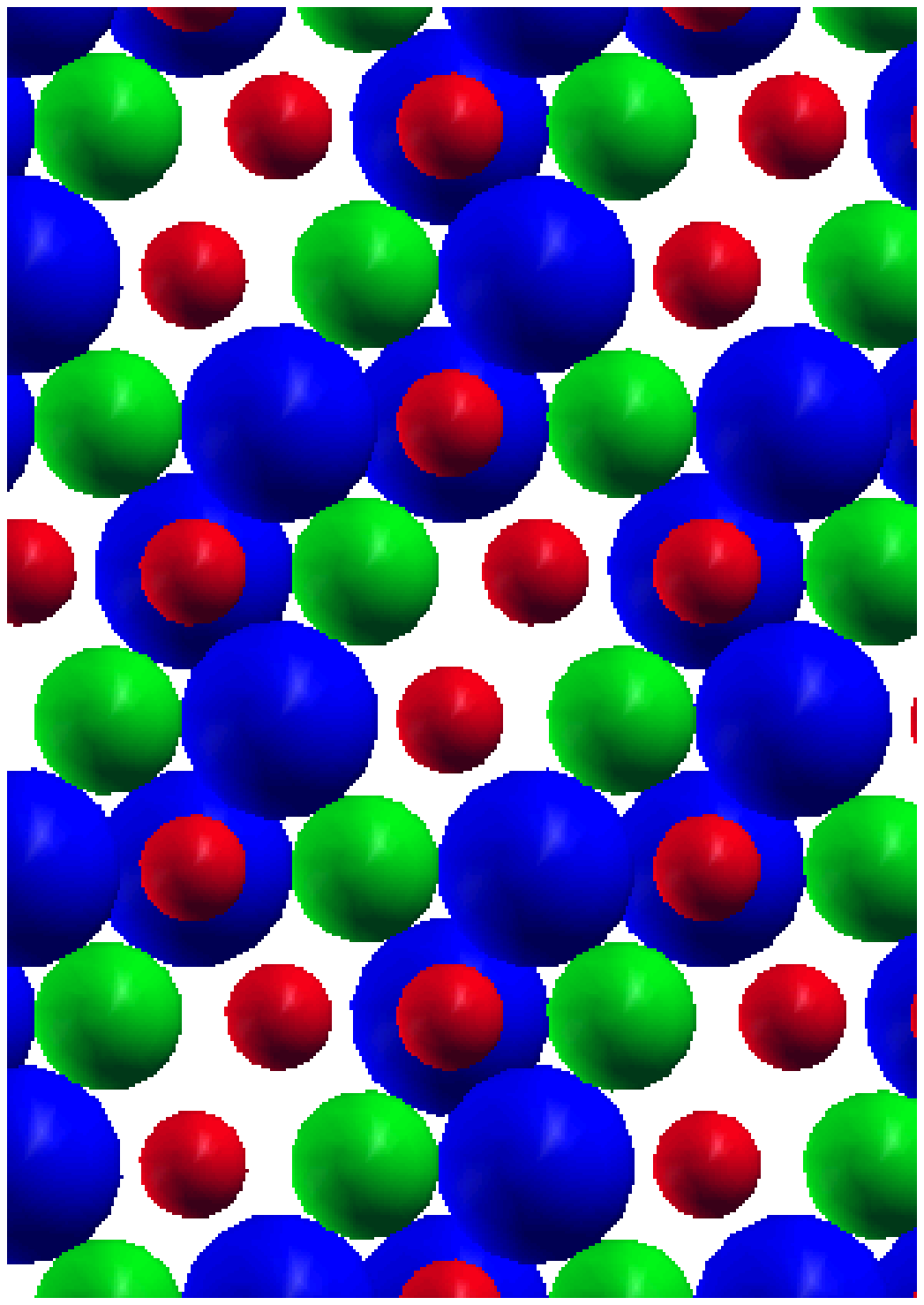}
\caption{(color online) Structure of two supercells, A (left)
and B (right). Na ions are
indicated by the large blue spheres, Co by the mid-size green spheres,
and O by the small red spheres.}
\label{struct}
\end{figure}

To investigate the effect of Na ion ordering, we focus on the $x=2/3$
composition and perform electronic structure calculations for
two supercells with different Na orderings and Co sites.
Specifically, we report results for two $\sqrt{3} \times \sqrt{3}$
cells with lattice parameters $a$=2.84 \AA, $c$=10.81 \AA.
Results are reported for the ideal structure,
with apical O height $z_{\rm O}$=0.0864 (Ref. \onlinecite{johannes}),
but with different Na orderings,
and for fully relaxed atomic positions.
Each supercell contains 6 Co and 12 Na sites in two layers.
The two supercells,
as shown in Fig. \ref{struct} are: (A) with one Na layer
on the Na$_{\rm Co}$ sites and the other on the Na$_h$ sites (both
with two Na atoms per layer); and (B) with both Na layers on the
Na$_h$ sites.

The calculations were done within the 
local density approximation with the general potential linearized
augmented planewave (LAPW) method including local orbitals. \cite{lapw,lo,cnote}
Well converged basis sets, of approximately 2600
LAPW functions plus local orbitals,
were used, with LAPW sphere
radii of 1.96 $a_0$ for the Na and Co and 1.52 $a_0$ for O.

\begin{figure}[tbp]
\epsfig{width=0.48\columnwidth,angle=0,file=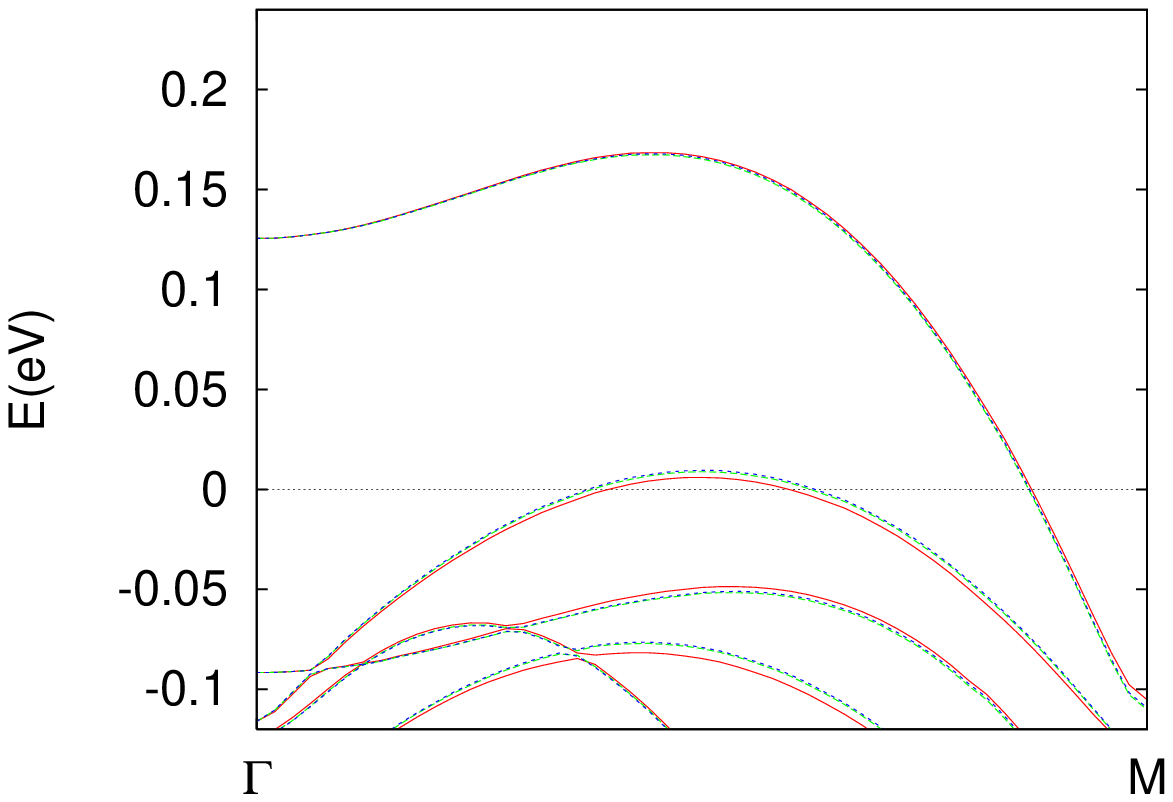}
\epsfig{width=0.48\columnwidth,angle=0,file=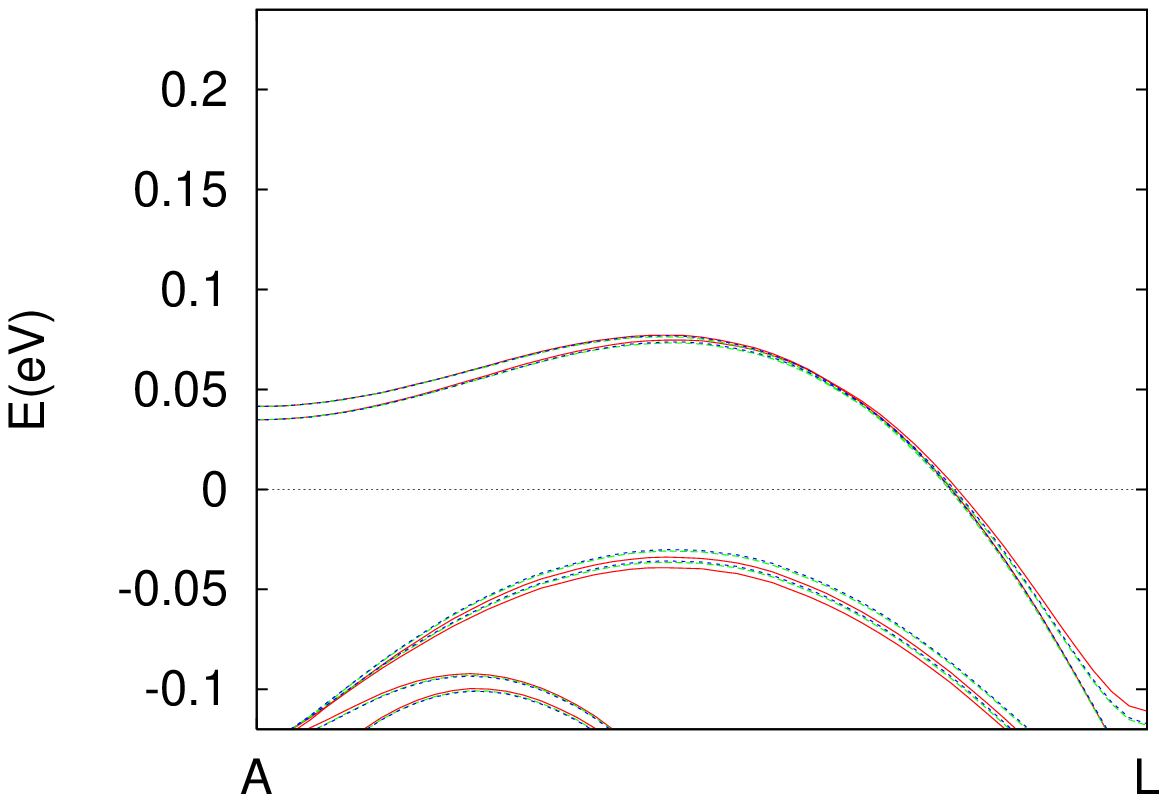}
\epsfig{width=0.48\columnwidth,angle=0,file=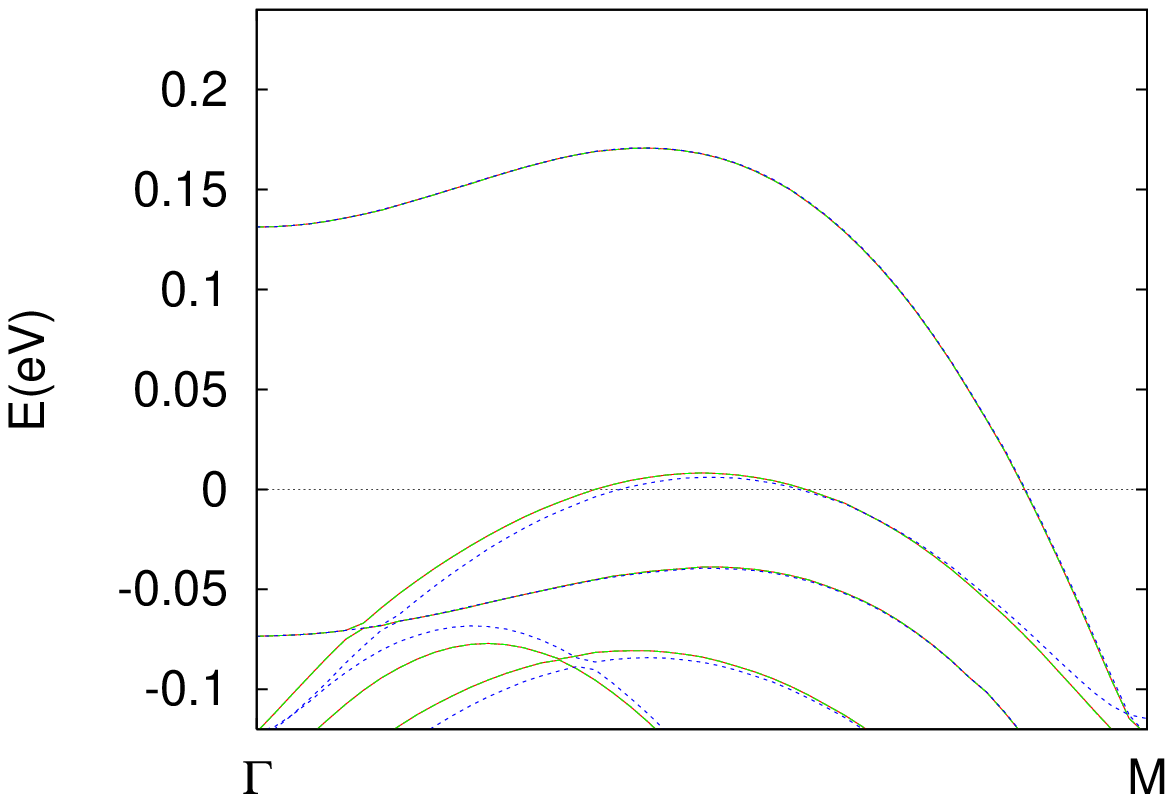}
\epsfig{width=0.48\columnwidth,angle=0,file=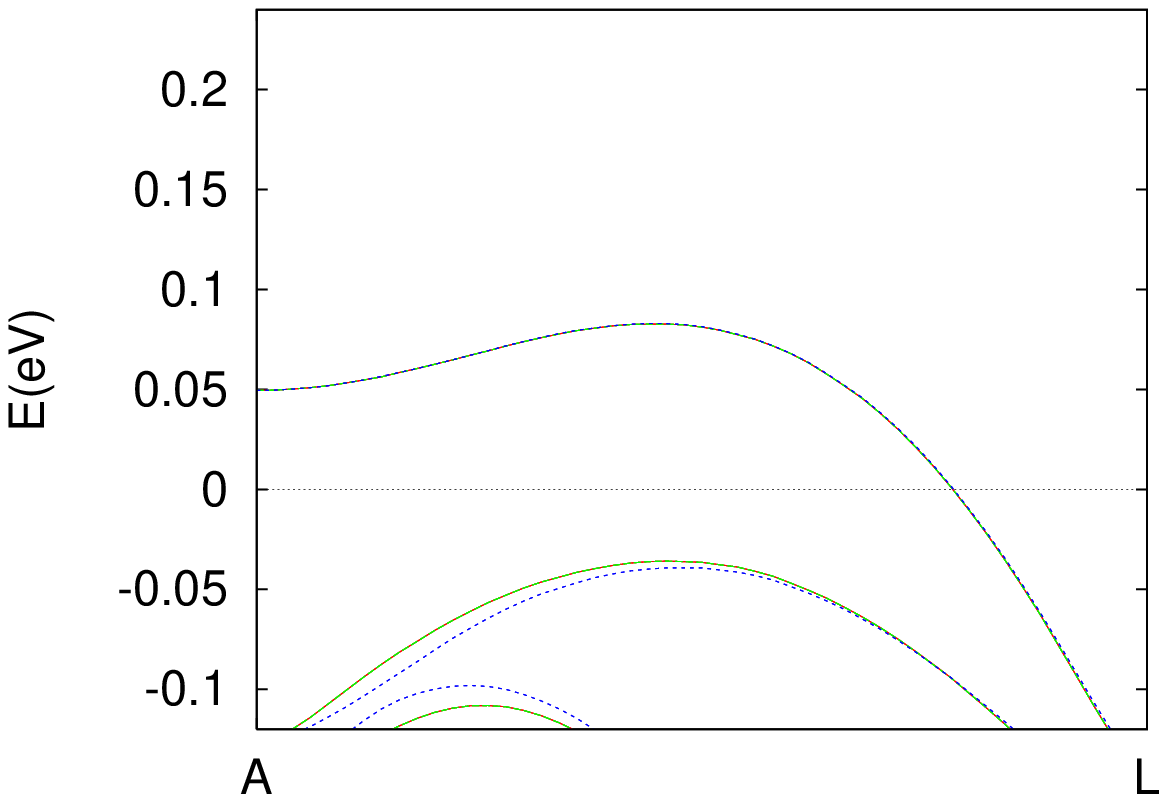}
\caption{(color online) Band dispersions along the $\Gamma$-$M$ and
$A$-$L$
directions for the two relaxed supercells: A (top) and B (bottom).
The different line types are for the three $\Gamma$-$M$ or
$A$-$L$ directions, separated by 120$^\circ$, which are non-equivalent
in the relaxed supercells.
}
\label{bands}
\end{figure}

The calculated band structures near the Fermi energy, $E_F$,
for the relaxed cells are shown
in Fig. \ref{bands}, along the supercell $\Gamma$-$M$ lines. Due
to the $\sqrt{3} \times \sqrt{3}$ supercell periodicity, these are
folded $\Gamma$-$K$ directions in the original hexagonal zone, and
therefore intersect the small pockets. The band structures obtained
for the two supercells are very similar to each other and to the virtual
crystal band structure. \cite{johannes}
In particular, at this doping level there is sufficient $k_z$
dispersion that only the outer of the two $a_g$ and $e_g'$ pairs
of bands produce Fermi surfaces in the $k_z=0$ plane, while both
$a_g$ and neither $e_g'$
bands produce Fermi surfaces at $k_z=1/2$.
This differs in detail from virtual crystal calculations, \cite{johannes}
where both $e_g'$ bands reach maxima above $E_F$ in the $k_z=1/2$
plane.
This difference in the $e_g'$, but not the $a_g$ topology, when using
real Na ions instead of the virtual crystal approximation is an
indication that variations in the Na potential due to disorder
could localize the $e_g'$ carriers.
In any case, the maximum of
the $a_g$ band is 0.17 eV above $E_F$, while the $e_g'$ maximum is only
0.01 eV above $E_F$. Besides the positions of the band maxima, there
is also a large difference in band velocity along
the $\Gamma$-$M$ line, $v(a_g)$=1.4 eV \AA,
$v(e_g')$=0.3 eV \AA.

The second indication of localization of the $e_g'$ carriers comes from
the Co and O 1$s$
core level positions. These measure the variation of the local
electrostatic potential in the supercells. Supercell A has two different
types of Co, the four Co coordinated by a Na in the prism immediately above
or below and the two without. In the unrelaxed cell those with Na
are 0.033 eV higher than the Co without. This variation is higher than
the energy difference between the
$e_g'$ band maximum and $E_F$.
Secondly, while supercell B has all Co equivalently
coordinated, the O sites differ in coordination in both supercells.
The variation in O core level position is 0.06 eV in both A and B.
This variation may be
significant because the bands are in fact hybridized Co-O
bands.
The variations in core level position are largely preserved in the relaxed
cells, and structural inhomogeneities in the Co-O bond lengths and
angles are introduced, which will also contribute to scattering.
The variation of Co 1$s$ levels is reduced to
0.029 eV in supercell A, {\em i.e.}
only slightly smaller than in the unrelaxed cell, and still larger than
the $e_g'$ band maximum relative to $E_F$, while the variation in the
O 1$s$ positions increases to 0.07 eV. \cite{posnote}
Thus the potential variations due to Na disorder are strong enough
to localize the $e_g$ states. \cite{palee}

The relaxation in supercell A is larger than in supercell B. This
is because the Co and nearby Na$_{\rm Co}$ ions are both positive and
move apart by 0.02 \AA, while the nearest neighbor Na-O distance contracts
by between 0.006 \AA$~$ to 0.01 \AA, depending on the site.
There are also distortions of the O cages.
Measured by the difference between the shortest and longest Co-O distance
for a given Co site, these range from 0.01 \AA (2/3 of the Co in B)
to 0.015 \AA$~$ (for the Co with Na$_{\rm Co}$
neighbors in A).

In any case, the results show that scattering due to Na disorder is
strong enough to localize the $e_g'$ carriers at $x=2/3$. The
actual position of the $e_g'$ band maximum relative to $E_F$ is
a function of doping, as is the Na distribution.
The linear size of the $e_g'$ Fermi surface, as predicted by band structure
calculations varies from zero at high $x$ to approximately 0.18\% of
the dimension of the zone at $x \sim 0.3$. \cite{johannes,johannes-nest}
At $x$=0.3, the $e_g'$ maximum lies 0.1 eV above $E_F$.
Observation of this Fermi surface would require a mean free path for $e_g'$
carriers of at least the inverse of this reciprocal space size.
At $x \sim 2/3$ this would imply $\ell_{eg'} >$ 50 \AA,
where $\ell_{eg'}$ is the mean free path for the heavy carriers on the
small Fermi surfaces. In metals with Fermi surfaces having similar
orbital character (here Co $t_{2g}$), the constant scattering
time approximation is reasonable. Then $\ell$=$v_F \tau$, where $\tau$
is a scattering time, $v_F$ is the Fermi velocity on a given section of
Fermi surface and $\ell$ is the mean free path for carriers on that
section. Taking the ratio of the velocities for the two sheets
along the $\Gamma$-$M$
line, the implication is that a mean free path $\ell_{ag} \sim 250$ \AA,
for the main Fermi surface would be needed in order for the small 
section to be seen. Note that the small sections are elliptical
in shape, elongated along
$\Gamma$-$M$, so the criterion in the tangential direction would be the same
as the velocity is larger, and the size smaller, in proportion.
Within kinetic transport theory, the conductivity is given by
$\sigma \propto N(E_F)v_F^2 \tau$, where $N(E_F)$ is the density of states
at $E_F$ and $v_F$ is the average Fermi velocity in the direction in
which the conductivity is measured. Thus the large difference in the
Fermi velocity between the large and small sections means that the
conductivity is dominated by the large section, regardless of whether
the small sections are present or not. Thus the measured resistivity
is, for practical purposes, governed by the
transport in the $a_g$ section. Depending on the doping level,
Na$_x$CoO$_2$ samples of the type used in photoemission have been reported
to have ratios of the resistivity at 300K to the residual resistivity
in the range of 20 - 30,
\cite{terasaki,yuyu}
and resistivity saturation at
or below $\sim 600K$.
These values are inconsistent with a 
mean free path at low temperature as long as 250 \AA.
We note that although the small sections become larger as $x$ is lowered,
they do so more slowly than in a rigid band picture and the large
ratio of the $v_F$ on the two surfaces is maintained.
\cite{johannes,johannes-nest}

Very recent high resolution photoemission work \cite{yang-2}
shows dispersions of the $a_g$ and $e_g'$ bands below $E_F$ that
have a noticeable anti-crossing along $\Gamma$-$K$, for which the
one electron $e_g'$ and $a_g$ bands have different symmetries and
would not mix.
The observed mixing is strong.
Furthermore, a non-dispersive spectral weight is seen just below
$E_F$ in the region where the $e_g'$ band is above $E_F$ in band
calculations. Both of these features are natural if scattering strong
enough to localize the $e_g'$ pockets is assumed.

To summarize we find that disorder in the Na layer of Na$_x$CoO$_2$
produces sufficiently strong potential variations to localize the
$e_g'$ Fermi surface pockets at least for high $x$. Other scattering
mechanisms and defects will enhance this tendency.
We argue that the non-observation of the small pockets and other
features seen in recent photoemission experiments can be
qualitatively understood within this framework, specifically
that the carriers associated with the small pocket are
localized due to scattering related to Na and other disorder.
In the hydrated superconducting compound, each Na is coordinated with
four H$_2$O molecules, is far from the CoO$_2$ sheets and the Na
is in a more
ordered pattern with respect to the Co atoms. One may speculate that
this leads to a sufficient reduction in the potential scattering due
to the Na, and that this could lead to the re-appearance of the $e_g'$
sections.

We are grateful for helpful discussions with H. Ding, M.Z. Hasan,
D. Mandrus, R. Jin, I.I. Mazin, M.D. Johannes and W.E. Pickett.
Research at ORNL
sponsored by the Division of Materials Sciences and Engineering,
Office of Basic Energy Sciences, U.S. Department of Energy,
under contract DE-AC05-00OR22725 with Oak Ridge National Laboratory,
managed and operated by UT-Battelle, LLC.
Work at UC Davis supported by DOE grant DE-FG03-01ER45876.


\begin{thebibliography}{100}
\bibitem{cnote}
Calculations were done using the WIEN2K code, \cite{wien}
and an independent code. These were cross-checked by calculations
of the core level positions using independent choices of parameters.
\end{thebibliography}
\end{document}